\documentclass[twocolumn]{aastex631}
\usepackage{multirow}
\usepackage{xspace}

\usepackage{amsmath}

\newcommand{\Msun}{\,{\rm M_\odot}}
\newcommand{\fedd}{\,{f_{\rm Edd}}}
\newcommand{\Mblack}{M_\bullet}
\newcommand{\Mstar}{M_\star}
\newcommand{\Ha}{\rm H\alpha}
\newcommand{\mmstar}{$\Mblack-\Mstar$\xspace}

\begin{document}

\title{The Redshift Evolution of the \mmstar Relation for JWST's Supermassive Black Holes at $z > 4$}

\author[0000-0001-9879-7780]{Fabio Pacucci}
\affiliation{Center for Astrophysics $\vert$ Harvard \& Smithsonian, Cambridge, MA 02138, USA}
\affiliation{Black Hole Initiative, Harvard University, Cambridge, MA 02138, USA}

\author[0000-0003-4330-287X]{Abraham Loeb}
\affiliation{Center for Astrophysics $\vert$ Harvard \& Smithsonian, Cambridge, MA 02138, USA}
\affiliation{Black Hole Initiative, Harvard University, Cambridge, MA 02138, USA}



\begin{abstract}
JWST has detected many overmassive galactic systems at $z > 4$, where the mass of the black hole, $\Mblack$, is $10-100$ times larger than expected from local relations, given the host's stellar mass, $\Mstar$. This Letter presents a model to describe these overmassive systems in the high-$z$ Universe. We suggest that the black hole mass is the main driver of high-$z$ star formation quenching. SMBHs globally impact their high-$z$ galaxies because their hosts are physically small, and the black holes have duty cycles close to unity at $z > 4$. In this regime, we assume that black hole mass growth is regulated by the quasar's output, while stellar mass growth is quenched by it and uncorrelated to the global properties of the host halo. We find that the ratio $\Mblack/\Mstar$ controls the average star formation efficiency: if $\Mblack/\Mstar > 8\times 10^{18} (n \Lambda/\fedd)[(\Omega_b M_h)/(\Omega_m M_\star) - 1]$, then the galaxy is unable to form stars efficiently. Once this ratio exceeds the threshold, a runaway process brings the originally overmassive system towards the local \mmstar relation. Furthermore, the \mmstar relation evolves with redshift as $\propto (1+z)^{5/2}$. At $z \sim 5$, we find an overmassive factor of $\sim 55$, in excellent agreement with current JWST data and the high-$z$ relation inferred from those. Extending the black hole horizon farther in redshift and lower in mass will test this model and improve our understanding of the early co-evolution of black holes and galaxies.
\end{abstract}


\section{Introduction} 
\label{sec:intro}
During the first year of operations of the James Webb Space Telescope (JWST), one of the most remarkable discoveries was the detection of a population of lower-mass ($10^6-10^8 \Msun$), lower-luminosity ($10^{44} - 10^{46} \, \rm erg \, s^{-1}$) supermassive black holes (SMBHs) at $z > 4$ \citep{Harikane_2023, Maiolino_2023_new, Ubler_2023, Stone_2023, Furtak_2023, Kokorev_2023, Yue_2023, Bogdan_2023}, reaching up to a redshift of $z=10.6$ with GN-z11 \citep{Maiolino_2023}. A comparison to the properties of the most distant quasar in the pre-JWST era, with a mass of $\Mblack = (1.6 \pm 0.4) \times 10^9 \Msun$ at $z = 7.6$, \citep{Wang_2021_quasar}, clarifies how much JWST has expanded our view on the early population of black holes, both upward in redshift and downward in mass.

The lower luminosity of these SMBHs allowed the detection of starlight from their hosts (see, e.g., \citealt{Ding_2022}) and estimate some of their properties, e.g., their stellar and dynamical mass and (gas) velocity dispersions. Some of these SMBHs were identified in the so-called ``little red dots'' \citep{Matthee_2023}, containing ``hidden little monsters'' \citep{Kocevski_2023}, which are low-luminosity, strikingly red objects. Recently, \cite{Greene_2023} used spectroscopy from the JWST/UNCOVER program to argue that $\sim 60\%$ of these objects are dust-reddened AGN: young galaxies hosting a low-luminosity SMBH at their center.

The discovery of a lower-luminosity population of SMBHs and their hosts' properties led to an additional, unexpected discovery. In the local Universe, well-known relations connect the mass of the central SMBHs with physical properties of their hosts (see, e.g., \citealt{Magorrian_1998, Ferrarese_Merritt_2000, Gebhardt_2000, Kormendy_Ho_2013}). For example, the \mmstar relation links the SMBH mass and the stellar mass of the host. \cite{Reines_Volonteri_2015} found that the mass of the central SMBH is $\sim 0.1 \%$ of the stellar mass of their hosts, with a scatter of $\sim 0.55$ dex, or a factor $\sim 3.5$.

A significant wealth of data from numerous JWST surveys indicates the detection of SMBHs at $z > 4$ that are $10-100$ times overmassive when compared to the stellar content of their hosts \citep{Harikane_2023, Maiolino_2023_new, Ubler_2023, Stone_2023, Furtak_2023, Kokorev_2023, Yue_2023}. The mass of these SMBHs is not $\sim 0.1 \%$ of the stellar mass of their hosts, but rather $1\%-10\%$, or even close to $\sim 100\%$ in some cases \citep{Bogdan_2023}.

A detailed statistical analysis of these data, with an MCMC algorithm that takes into account observational biases (e.g., see \citealt{Lauer_2007}), finds that this population of lower-mass SMBHs at $z > 4$ violate the \mmstar relation at  $> 3 \sigma$ \citep{Pacucci_2023_JWST}. 
Interestingly, \cite{Maiolino_2023_new} notes that while the SMBHs are overmassive with respect to the \mmstar relation, other scaling relations, such as the $\Mblack-\sigma$ and the $\Mblack - M_{\rm dyn}$ relations (with the velocity dispersion and the dynamical mass, respectively), hold at $4 < z < 7$. 
Altogether, recent JWST data suggests that the $\Mblack-\sigma$ and the $\Mblack - M_{\rm dyn}$ relations are ``fundamental and universal'' because they are powered by the depth of the gravitational potential well generated by the central SMBH.
Instead, the \mmstar relation could evolve with redshift.

The $\Mblack-\sigma$ and the $\Mblack - M_{\rm dyn}$ relations are linked, so it is not surprising that once one holds, the other follows. Instead, the host's stellar mass is measured independently. 
Despite significant uncertainties affecting stellar mass and black hole mass measurements, \cite{Pacucci_2023_JWST} find that, unless most overmassive SMBHs found so far are characterized by errors of a factor $\sim 60$ in their black hole mass or their stellar mass all in the same direction (i.e., all increasing $M_\star$ or decreasing $\Mblack$), this result holds. Note that typical reported errors, at $1\sigma$, are of a factor $\sim 3$ in $\Mblack$ and a factor $\sim 4$ in $M_\star$ (see, e.g., \citealt{Maiolino_2023_new}).

Theoretical predictions, dating back 20 years, suggest that scaling relations evolve with redshift. For instance, \cite{Wyithe_Loeb_2003} argued that the ratio $\Mblack/M_\star$ should scale as $\propto (1+z)^{3/2}$, due to self-regulation via quasar \citep{Silk_1998} and supernova feedback. Quasar activity is efficient in quenching star formation via the effect of strong outflows or by heating the gas (e.g., \citealt{Fabian_2012, Heckman_2014, King_Pounds_2015}), although there is at least one example of a black hole triggering star formation in a dwarf galaxy \citep{Schutte_2022}.

Other works have studied the redshift evolution of scaling relations using numerical simulations (e.g., \citealt{Robertson_2006, DiMatteo_2008, Sijacki_2015}), semi-analytic models (e.g., \citealt{Malbon_2007, Kisaka_2010}), observations (e.g., \citealt{Peng_2006, Decarli_2010, Merloni_2010, Trakhtenbrot_2010, Bennert_2011}), or combinations of those (e.g., \citealt{Booth_2011}), especially at $z \lesssim 2$. More recently, \cite{Caplar_2018} developed a phenomenological model, based on observations, and found that the ratio $\Mblack/M_\star$ scales as $\propto (1+z)^{1.5}$ at $z < 2$, in agreement with \cite{Wyithe_Loeb_2003}. Furthermore, using simple but effective arguments on the density and dust content of the high-$z$ population of galaxies detected by JWST, very recently \cite{Inayoshi_2024} proposed a high-$z$ \mmstar that is in complete agreement with ours.

In this Letter, we present a model that explains the evolution at $z > 4$ of the \mmstar relation for SMBHs in JWST data. Furthermore, we develop a condition on the ratio $\Mblack/M_\star$ to probe whether the quasar feedback stunts star formation. The model makes predictions that can be tested with future JWST data.

\section{Model Principles}
\label{sec:theory}
We start with the principles of our model for high-$z$ overmassive systems.
The model we present is valid only in the high-$z$ Universe, where the typical growth time for black holes, $t_g$, is similar to the age of the Universe: $t_g \sim t_{age}$.

The basis of our model is that the \textit{black hole mass is the primary parameter that controls high-$z$ star formation quenching}. This premise is supported by recent analyses of JWST/CEERS data with cosmological simulations (Illustris TNG and EAGLE), showing that high-$z$ galaxy quenching is primarily regulated by the mass of the SMBH \citep{Piotrowska_2022, Bluck_2023}. Previous cosmological simulations already showed that star formation quenching should scale with energy input from the central SMBH over the entire lifetime of the galaxy, which is proportional to the black hole mass \citep{Terrazas_2020, Bluck_2023}.
Active SMBHs in $z > 4$ galaxies discovered by JWST are effective at quenching star formation for at least two reasons.

First, \textit{high-$z$ galaxies are physically small}; the ionized bubble generated by active SMBHs likely extends to the entire galaxy. Typical physical sizes of galaxies detected by JWST at $z=7-9$ are characterized by effective radii of $80 \, \mathrm{pc} < r_e < 300 \, \mathrm{pc}$, with a mean value of $r_e \sim 150$ pc \citep{Baggen_2023}. Recently, \cite{Baldwin_2024} estimated the size of GN-z11 as $150 \pm 25$ pc. This typical physical size has to be compared with the radius of the ionization bubble created by a SMBH accreting at its Eddington rate. For a $\sim 10^7 \Msun$ SMBH, as typically found in these overmassive systems \citep{Pacucci_2023_JWST}, this radius can extend as much as $\sim 700$ kpc (see, e.g., \citealt{Cen_Haiman_2000, Madau_Rees_2000, White_2003}). Regions of high-density gas presumably present in the high-$z$ galaxies could effectively shield the radiation from the SMBH and still allow localized star formation; this would, however, be ineffective in generating large-scale star formation. Hence, SMBHs were effective in quenching star formation because they had a global impact on the entire host. Note that previous studies (e.g., \citealt{Chen_2020_radius}) already highlighted the importance of the radius of star-forming galaxies in determining the growth of their SMBHs.

\begin{figure*}%
    \centering
\includegraphics[angle=0,width=0.49\textwidth]{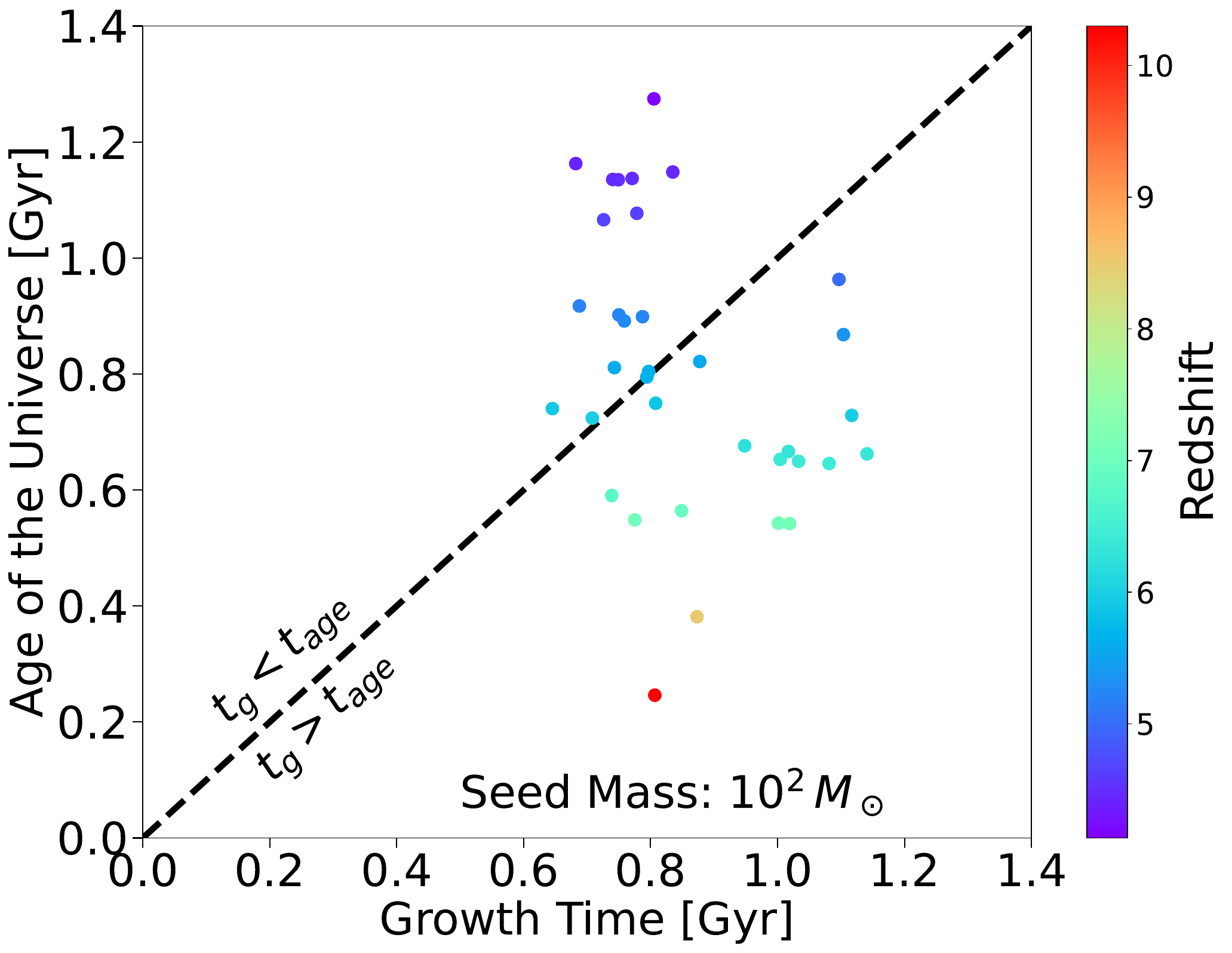} \hfill \includegraphics[angle=0,width=0.49\textwidth]{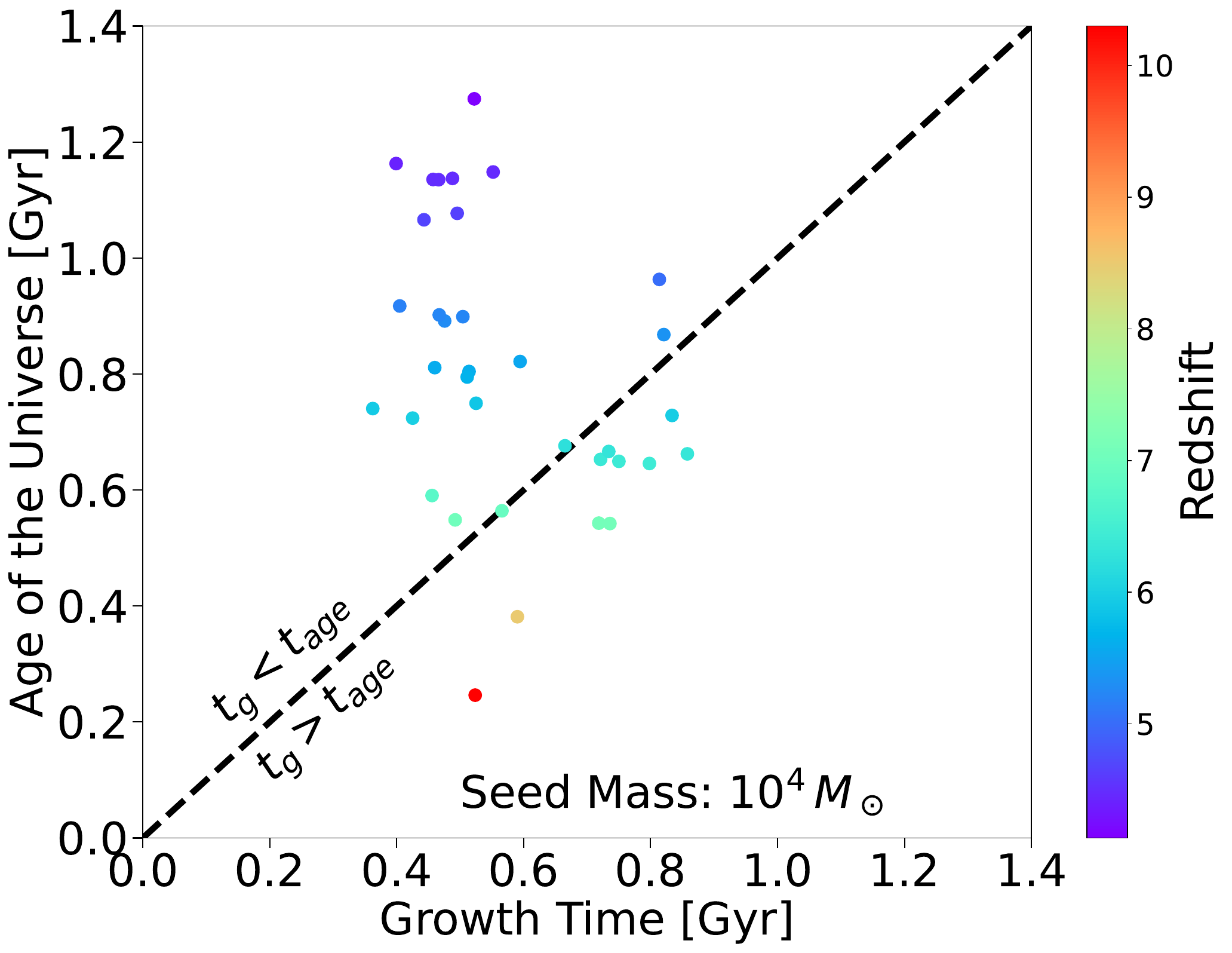}
    \caption{\textbf{Left panel:} Comparison between the growth time (assuming a light seed of $100 \Msun$) and the age of the Universe at detection, for the overmassive systems detected by JWST thus far. The dashed line indicates where the growth time equals the Universe's age at that detection redshift. The data points are colored according to their detection redshift, shown in the color bar. \textbf{Right panel:} same as the left panel, but the growth time assumes a heavy seed of $10^4 \Msun$.}
    \label{fig:ages}%
\end{figure*}

Second, \textit{in the high-$z$ Universe, the growth time of black holes is comparable to the age of the Universe.} Hence, central SMBHs at a given redshift $z$ are active for a time comparable to the Hubble time $t_{age}(z)$, if $z > 4$.
Figure \ref{fig:ages} shows, for the high-$z$ overmassive systems detected thus far by JWST, a comparison between the age of the Universe (at detection) and growth time. This latter time is calculated assuming a continuous growth from a seeding redshift of $z = 25$ (see, e.g., \citealt{BL01}) at the Eddington rate, assuming a light seed of $100 \Msun$ (left panel) or a heavy seed of $10^4 \Msun$ (right panel). At higher redshift, the age of the Universe is comparable to, or even shorter than, the growth time (for those two specific seeding scenarios); hence, the ``activity duty cycle'' for those specific SMBHs has to be close to unity.
An example to clarify our point. At the median redshift $z=5$ \citep{Pacucci_2023_JWST}, the age of the Universe is $\sim 1.1$ Gyr. To reach a typical black hole mass of $10^8 \Msun$, accreting at the Eddington rate, a light seed of $10^2 \Msun$ would take $63\%$ of the age of the Universe (i.e., $692$ Myr); a heavy seed of $10^5 \Msun$ would take $31\%$ of the age of the Universe (i.e., $346$ Myr). These black holes had to be ``on'' for a fraction of the age of the Universe close to unity, independent of the seeding mechanism.
The seed mass is an important parameter but affects the growth time only logarithmically. In the example above, three orders of magnitude change in the seed mass affects the growth time only by a factor $\sim 2$. Such factors can be crucial in the very high-$z$ Universe, and to form extremely massive SMBHs. However, in the present study, we are describing SMBHs at typical redshifts of $\sim 5$. Hence, we argue that our model principles do not depend on the particular flavor of the black hole seed chosen.

Because their duty cycle is close to unity, these SMBHs are constantly injecting energy into the primeval galaxy and heating the cold gas necessary to produce stars. Once the age of the Universe is $ \gtrsim 1$ Gyr, the SMBH is active only for a fraction of the Hubble time and stars can then form from cold molecular gas, which becomes widely available in the galaxy.
This hypothesis is further confirmed by a recent analysis of $4.5 <  z < 12 $ galaxies in the JWST/CEERS survey \citep{Cole_2023}, showing a higher variability of star formation activity at high redshift. Stars form primarily in short periods of starburst activity, with star-forming duty cycles of only $20\%$ at $z \sim 9$, and $40\%$ at $z \sim 5$. The study also suggests a smoother star formation activity at $z < 4.5$, when the age of the Universe is $>1 \, \rm Gyr$ and the quasar duty cycle drops significantly below unity.

\subsection{Assumptions on the Growth of $\Mblack$ and $\Mstar$}
The two assumptions in our model for high-$z$ overmassive systems are the following ($v_c$ is the circular velocity of the galactic halo, see \citealt{BL01}):
\begin{enumerate}
    \item Black hole mass growth is regulated by the quasar output. This leads to the scaling $\Mblack \propto v_c^5$.
    \item Stellar mass growth is quenched by the quasar output and uncorrelated with $v_c$: $\Mstar \not \propto v_c$.
\end{enumerate}

The first scaling is easily demonstrated as follows (see the same derivation in \citealt{Wyithe_Loeb_2003}).
Assume that the central SMBH of mass $\Mblack$ is emitting energy at a fraction $\eta$ of the Eddington luminosity, $L_{\rm Edd}$, with $L_{\rm Edd} \propto \Mblack$. Let us further assume that a fraction ${\cal F}$ of $\eta L_{\rm Edd}$ is trapped by the gas within the galaxy. The self-regulation hypothesis predicts that the growth of the central SMBH shuts off when the total luminosity output of the SMBH, absorbed by the gas over a dynamical time $t_{\rm dyn}$, is equal to the binding energy of the host halo of total mass $M_h$:
\begin{equation}
    \eta L_{\rm Edd} {\cal F} = \frac{1}{2} \frac{\Omega_b M_h v_c^2}{\Omega_m t_{\rm dyn}} \, ,
    \label{eq:regulation_qso}
\end{equation}
where $\Omega_b/\Omega_m$ is the baryon fraction.

From \cite{BL01}, we derive the dependence of the circular velocity of stars with respect to the halo mass $M_h$ and the redshift $z$:

\begin{equation}
    v_c=245\left(\frac{M_{h}}{10^{12} M_{\odot}}\right)^{1 / 3}[\xi(z)]^{1 / 6}\left(\frac{1+z}{3}\right)^{1 / 2} \mathrm{~km} \mathrm{~s}^{-1},
    \label{eq:vc}
\end{equation}
where the factor $\xi(z)$ is defined as:
$$
\begin{gathered}
\xi \equiv \frac{\Omega_m}{\Omega_m^z} \frac{\Delta_c}{18 \pi^2}, \\
\Omega_m^z \equiv\left[1+\left(\frac{\Omega_{\Lambda}}{\Omega_m}\right)(1+z)^{-3}\right]^{-1}, \\
\Delta_c=18 \pi^2+82 d-39 d^2, \\ 
d=\Omega_m^z-1
\end{gathered}
$$
As $M_h \propto v_c^3$, our previous self-regulation equation implies that $\Mblack \propto v_c^5$.

Regarding the second assumption, it is essential to note that the circular velocity $v_c$ depends on the total mass of the halo, not on its stellar mass.
We argue that the total halo mass corresponding to a given circular velocity $v_c$ is in place. However, it is not forming stars efficiently because their growth is inhibited by high-duty cycle quasar activity in a small-size galaxy.

This second assumption can be tested experimentally. From JWST observations at redshift $4 < z < 7$ (see, e.g., \citealt{Maiolino_2023_new}), there is no relation between the stellar mass of the host and the measured velocity dispersion of the galaxy. While the $\sigma$ values vary in the \cite{Maiolino_2023_new} dataset in a range of $0.3$ dex, the stellar masses vary for $2.5$ dex. A Pearson's (2-tailed) correlation test yields no correlation (p-value $\sim 0.03$) at $5\%$ significance.
Hence, $M_\star \propto \sigma^0 \propto v_c^0$.
At these redshifts and for these stellar masses, there is no indication that the stellar mass growth is regulated by either supernova or the quasar feedback. In this regard, our treatment is fundamentally different from \cite{Wyithe_Loeb_2003}.

\section{Results}
\label{sec:results}
Next, we derive our results: a condition on the ratio $\Mblack/M_\star$ for the average star formation rate to be effectively quenched at high-$z$ (Sec. \ref{sec:condition}) and a prediction for the redshift evolution of the \mmstar relation at $z > 4$ (Sec. \ref{sec:evolution}).

\subsection{A Condition for Star Formation Quenching}
\label{sec:condition}

We have developed a theoretical condition on the ratio $\Mblack/M_\star$ to understand if quasar feedback is efficient in quenching star formation and to which extent. We then test this hypothesis with the $35$ overmassive systems discovered by JWST at $z > 4$ \citep{Harikane_2023, Maiolino_2023_new, Ubler_2023, Stone_2023, Furtak_2023, Kokorev_2023, Yue_2023, Bogdan_2023}.

Before proceeding, we note that galaxies meeting the condition for star formation quenching developed here are \textit{not prevented} from forming stars \textit{altogether}, or even at the observation time. After all, the $35$ overmassive systems studied here contain $10^8-10^{11} \Msun$ in stars, which must have formed at some point. Likely, the existing stellar masses were formed when the black hole mass was small, and the quasar feedback was weak.
Our condition on the ratio $\Mblack/M_\star$ prevents star formation from being efficient over the entire lifetime of the galactic system, up to detection. In other words, the criterion we developed applies to the \textit{time average of the star formation rate}, not to its instantaneous value. Despite this note, it is reassuring to report that out of the $35$ overmassive systems studied here, only $3$ display large far-infrared luminosities, which may indicate ongoing star formation \citep{Stone_2023}.

We assume that star formation is quenched once the SMBH injects sufficient energy into the system to raise the temperature above the virial one. 
Quasar feedback in the form of heating, not mechanical outflows, is thus responsible for quenching the average star formation efficiency; a multitude of studies have investigated the interplay between quasar feedback (both thermal and mechanical) and the formation of stars, both in the local and the high-$z$ Universe (see, e.g., \citealt{Silk_1998, King_2003, Hickox_2009, Cattaneo_2009, Inayoshi_2014, Weinberger_2017}).
Recently, \cite{Gelli_2023} used a similar formalism to argue that supernova feedback fails to quench star formation in high-$z$ galaxies. This finding supports our model principles detailed in Sec. \ref{sec:theory}.

A simple model to describe the energetics of a primordial galaxy includes $\cal H$ and $\cal C$: the rate of energy injection (heating) and the rate of energy subtraction (cooling).

The power injected into the system, for a SMBH accreting at Eddington ratio $\fedd$ (defined as the ratio between the actual accretion rate and the Eddington rate), is
\begin{equation}
    {\cal H} = \fedd \frac{4 \pi G \Mblack m_p c}{\sigma_T} \, ,
\end{equation}
where $G$ is the gravitational constant, $m_p$ is the proton mass, $c$ is the speed of light, and $\sigma_T$ is the Thomson cross section.

The energy of a gas characterized by pure translational kinetic energy, at the virial temperature $T_{\rm vir}$, is $E = \langle 3/2\rangle N k_B T_{\rm vir}$, where $N$ is the total number of particles and $k_B$ is the Boltzmann constant.
Expressing the cooling time as $t_{\rm cool} \simeq 3k_B T_{\rm vir}/(\Lambda n)$, where $\Lambda$ is the cooling function and $n$ is the gas number density (see, e.g., \citealt{Rees_Ostriker_1977, BL01}), we can express the cooling rate as:
\begin{equation}
    {\cal C} = \frac{1}{2} \frac{n\Lambda}{\mu m_p} M_g \, .
\end{equation}
Here, $\Lambda(T_{\rm vir}, Z)$ is the cooling function in terms of the virial temperature and metallicity, $\mu = 0.6$ is the mean molecular weight for ionized gas, and $M_g$ is the gas mass. This last term can be expressed as the baryon mass of the halo minus the mass in stars: $M_g = (\Omega_b/\Omega_m)M_h - \Mstar$.
Note that the cooling function $\Lambda$ has units $\rm erg \, s^{-1} \,  cm^{3}$.

The condition that feedback heats the gas mass above its virial temperature in the small, high-$z$ galaxy is: ${\cal H} > {\cal C}$.
This condition can be expressed in terms of the ratio $\Mblack/M_\star$ as follows:
\begin{equation}
    \frac{\Mblack}{\Mstar} > \frac{1}{8\pi} \frac{n \Lambda \sigma_T}{G c \mu m_p^2 \fedd} \left( \frac{\Omega_b}{\Omega_m}\frac{M_h}{\Mstar} - 1 \right) \, .
\end{equation}
Expressing the constants in numerical form (with units equal to the reciprocal of $\rm erg \, s^{-1}$), this translates into:
\begin{equation}
    \frac{\Mblack}{\Mstar} > 8 \times 10^{18} \frac{n \Lambda}{\fedd} \left( \frac{\Omega_b}{\Omega_m}\frac{M_h}{\Mstar} - 1 \right) \, .
\end{equation}
We adopt $\fedd = 1$ for our calculations because overmassive systems discovered by JWST are estimated to be accreting at rates $0.1 < \fedd < 5$ \citep{Harikane_2023, Maiolino_2023_new}.
In particular, the Eddington ratio distribution of sources described in those two studies is skewed towards higher values, and well described by the statistics $\fedd = 0.9^{+1.4}_{-0.3}$, where $0.9$ is the mean, and the upper and lower bounds are derived from the interquartile ranges. In this simple model, we implicitly assume that the fraction of the energy emitted by the quasar that is retained by the gas, $F_q$, is constant and does not depend on the stellar mass (see, e.g., \citealt{Ferrarese_2002, Wyithe_Loeb_2003, Begelman_2004}). The role of $F_q$ is similar to, and degenerate with, the Eddington ratio. In terms of heating, a higher fraction of energy retained by the gas would play the same role as a higher Eddington ratio. Current data do not warrant a more complex interdependence between $F_q$ and $\Mstar$.

Furthermore, we use the median cooling function for the gas metallicity range $ 0.1 < Z/Z_\odot < 0.3$ and the mean gas number density ($\sim 0.5 \, \mathrm{cm^{-3}}$) calculated for simulated galaxies in the redshift range $5 < z < 10$ by \cite{Robinson_2022}. The metallicity range used is justified by a recent study with JWST of $z\sim 6$ galaxies with masses $\sim 10^{10} \Msun$, showing typical values $12 + \log(\mathrm{O/H}) \sim 8.2$, which is $\approx 25\%$ of the solar value \citep{Nakajima_2023}.
Finally, we assume the values of the cosmological parameters from \cite{Planck_2018} and the halo mass to stellar mass ratio from \cite{Behroozi_2019}.

\begin{figure*}%
    \centering
\includegraphics[angle=0,width=0.80\textwidth]{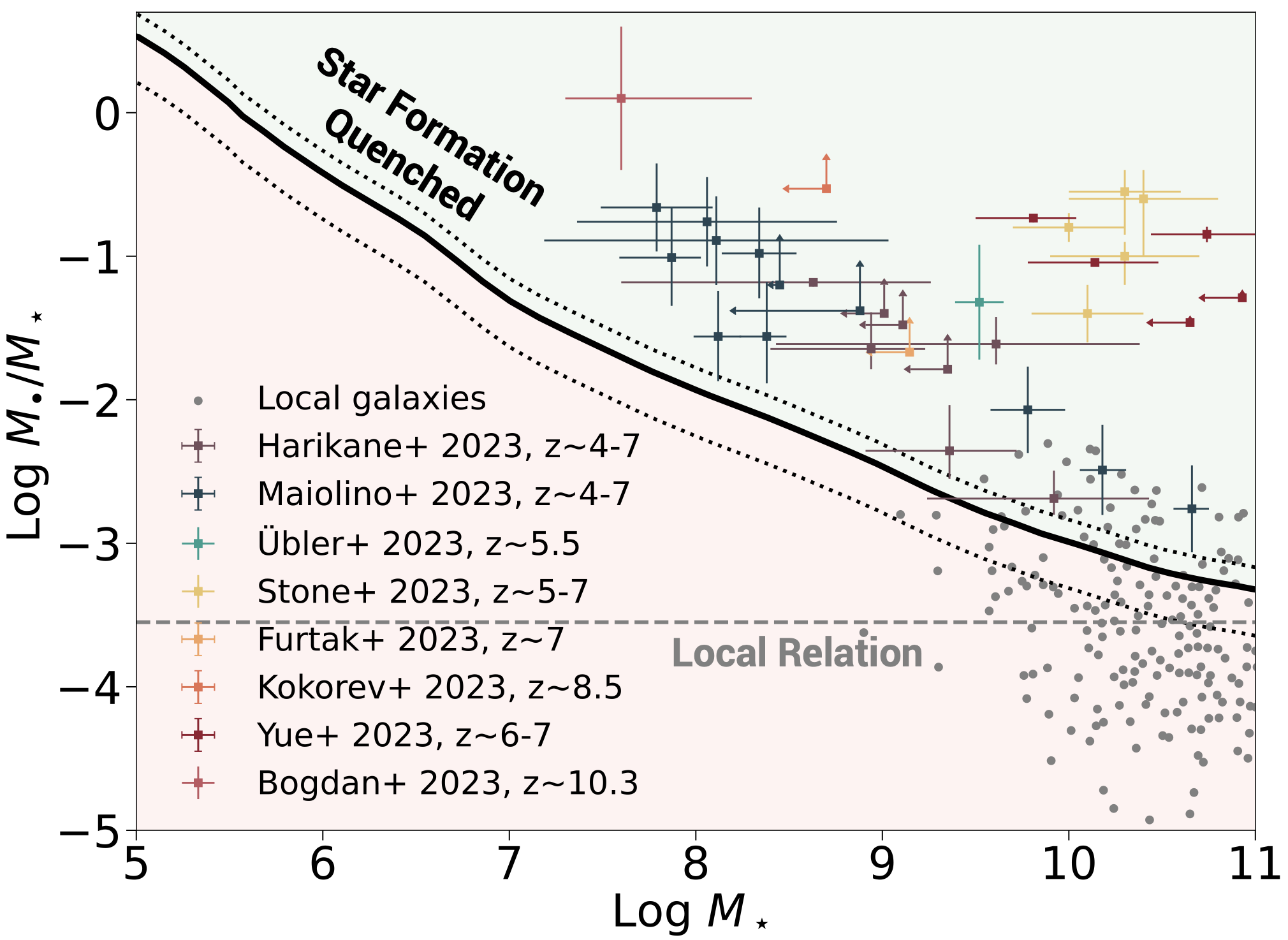} \hfill
    \caption{Condition on the ratio $\Mblack/M_\star$ for quasar feedback to suppress the \textit{average} star formation efficiency. Active galaxies that reside in the green area, with a ratio $\Mblack/M_\star$ above the threshold (indicated with a black line), experience quasar activity that increases the gas temperature above the virial value. Colored symbols indicate overmassive systems discovered by JWST at $z > 4$, as the legend indicates. Gray symbols indicate local galaxies on the \mmstar relation from \cite{Reines_Volonteri_2015}, whose ratio $\Mblack/M_\star$ is shown as a dashed line. The dotted lines indicate how the threshold ratio is affected by the range of variability of Eddington ratios in the high-$z$ sample considered; this is estimated with interquartile ranges, to properly describe skewness.}
    \label{fig:condition}%
\end{figure*}

In Fig. \ref{fig:condition}, we show the condition on the ratio $\Mblack/M_\star$. 
First, we note that \textit{for large stellar masses} ($\Mstar > 10^{11} \Msun$) \textit{the threshold ratio tends to the local one: $\log_{10}(\Mblack/\Mstar) \sim -3$}. This indicates that quasar quenching of overmassive systems at high-$z$ leads naturally to a ratio similar to the one implicit in the local relation.  
Central SMBHs grow until feedback self-regulates it, or the available gas runs out. Then, when the quasar's duty cycle drops below unity, efficient star formation can resume; mergers with other galaxies also bring additional mass in stars. Eventually, stellar mass growth by in-situ formation and mergers pushes the system below the threshold and towards the local \mmstar relation.

Note that \textit{this threshold can only be crossed once in the downward direction}. If a SMBH is overmassive, it will decrease the average star formation efficiency until it shuts off. Once stars begin to form again, the system will move downward and eventually cross the threshold. At that point, star formation is not quenched anymore; a runaway process occurs that pushes the system more into the star-forming region. This process ends with the system close to the local \mmstar relation.

Figure \ref{fig:condition} shows the location of the aforementioned $35$ overmassive systems discovered by JWST at $z > 4$. All these systems are either well inside the area where star formation is quenched or close to the threshold value. Typically, higher redshift systems (i.e., with $ z > 5$, see the ones by \citealt{Kokorev_2023} and \citealt{Bogdan_2023}) are deeper into the quenching regime than lower redshift ones, with $z \sim 4$. The only system whose location is marginally compatible with the threshold, possibly indicating that the galaxy is about to restart efficient star formation, is CEERS $01665$, at $z = 4.483$ \citep{Harikane_2023}.

In Fig. \ref{fig:condition} we also show a sample of local $z \sim 0$ galaxies from \cite{Reines_Volonteri_2015}, which are used to infer the local \mmstar relation. Although the threshold ratio $\Mblack/M_\star$ is computed for high-$z$ systems and not necessarily valid in the local Universe, it is reassuring to see that most of the local galaxies on the \mmstar relation reside well into the regime where star formation is active, or close to the boundary. This fact further suggests that high-$z$ overmassive systems migrate towards the local \mmstar relation by crossing the threshold once. Some of the $z \sim 4$ overmassive systems share their locus in the diagram with these local galaxies.

\subsection{The Redshift Evolution of the \mmstar Relation}
\label{sec:evolution}
\begin{figure}%
    \centering
\includegraphics[angle=0,width=0.45\textwidth]{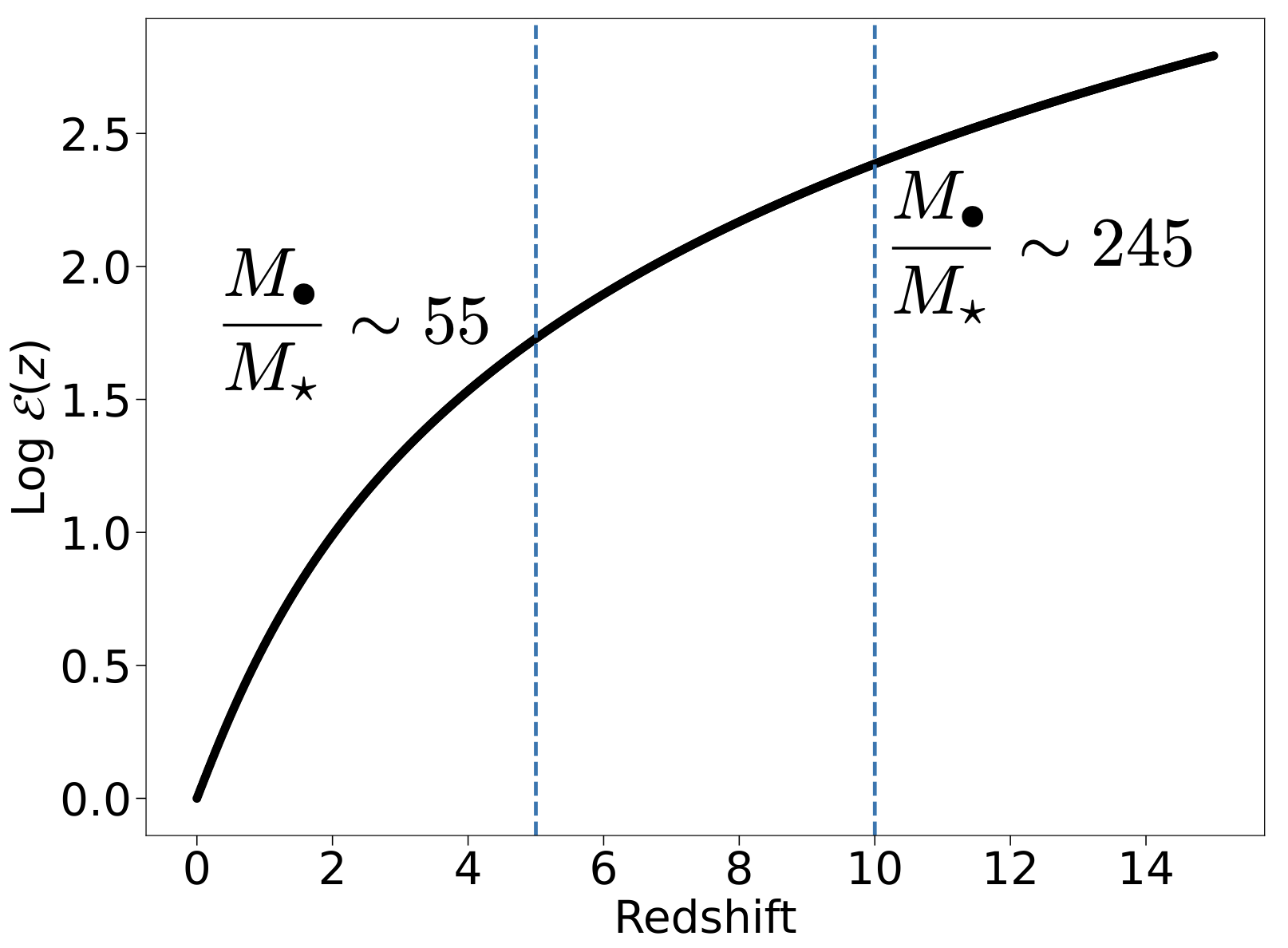} \hfill
    \caption{Value of the logarithm in base 10 of ${\cal E}(z)$ for $ 0 < z < 15$. The values for $z = 5$ and $z = 10$ are marked and indicated.}
    \label{fig:ratio}%
\end{figure}

We now derive a function to describe the redshift evolution of the \mmstar relation. Based on the two principles described in Sec. \ref{sec:theory},  we obtained the following scaling for $\Mblack$ and $\Mstar$ as a function of the circular velocity of the host halo: $\Mblack \propto v_c^{5}$ and $M_\star \propto v_c^0$.
Hence, $\Mblack/M_\star \propto v_c^{5}$.
Given the scaling of the circular velocity with redshift (Eq. \ref{eq:vc}), we obtain:
\begin{equation}
    \frac{\Mblack}{M_\star} \propto \xi(z)^{5/6} (1+z)^{5/2} \, .
\end{equation}
Note that $\xi(z)$ is a weakly varying function of the redshift; the main scaling is with the term $(1+z)^{5/2}$.
We define a redshift evolution function ${\cal E}(z)$:
\begin{equation}
    {\cal E}(z) = \frac{\xi(z)^{5/6} (1+z)^{5/2}}{\xi(0)^{5/6}} \, ,
    \label{eq:z_evolution}
\end{equation}
and note that ${\cal E}(z)$ indicates how much SMBHs at redshift $z$ are overmassive when compared to what is expected from local ($z=0$) relations.
The value of ${\cal E}(z)$ for $ 0 < z < 15$ is shown in Fig. \ref{fig:ratio}.

\begin{figure*}%
    \centering
\includegraphics[angle=0,width=0.80\textwidth]{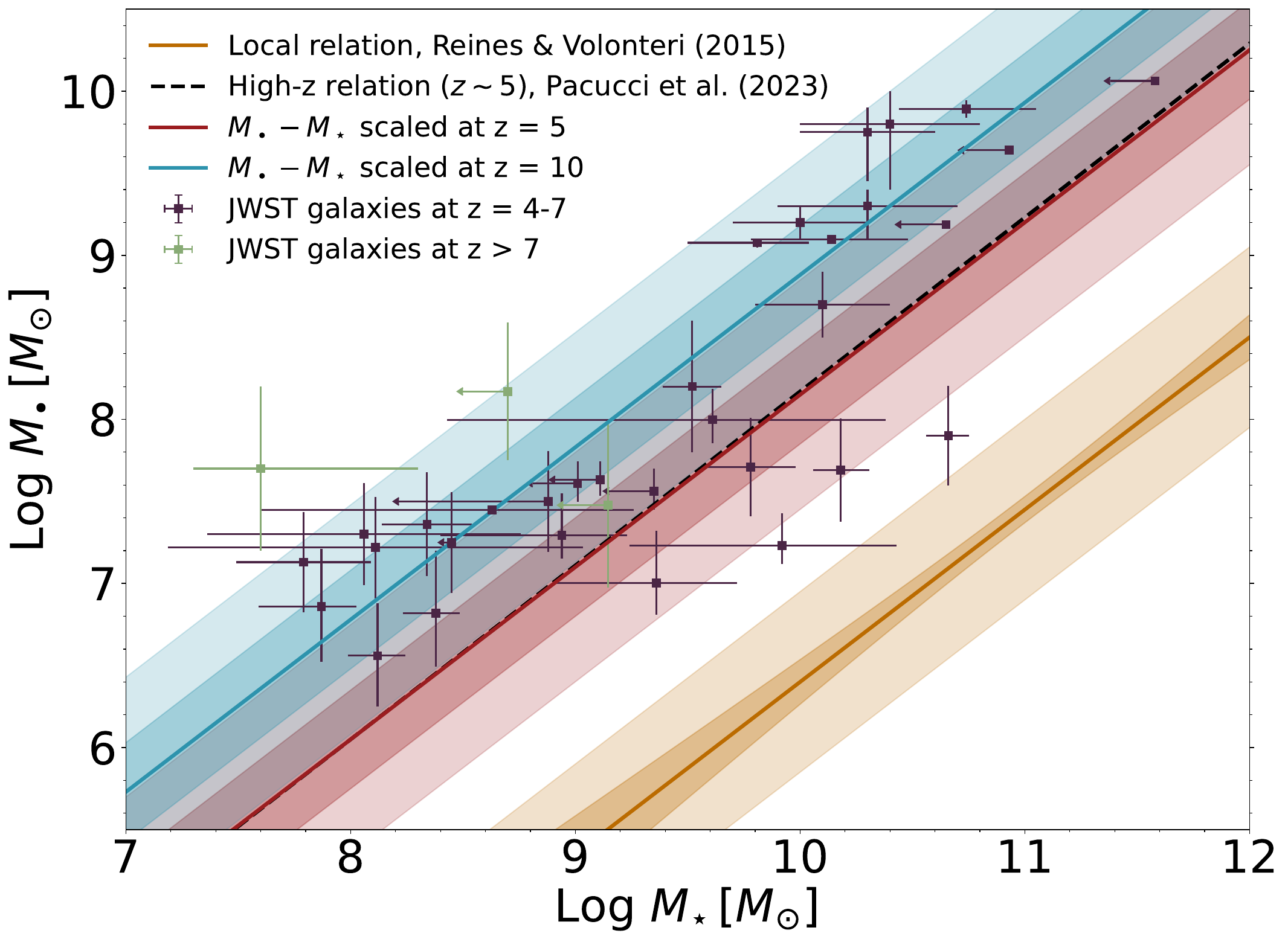} \hfill
    \caption{The \mmstar plane is populated with the overmassive systems discovered by JWST at $z > 4$ (categorized into two groups: $4<z< 7$ and $z > 7$). The local relation \citep{Reines_Volonteri_2015} is shown in yellow and scaled up at $z=5$ (red) and $z=10$ (blue), according to Eq. \ref{eq:z_evolution}. The dashed, black line indicates the high-$z$ relation inferred from JWST data by \cite{Pacucci_2023_JWST}. Our model for the redshift evolution of the \mmstar relation predicts the trend remarkably well.}
    \label{fig:mmstar}%
\end{figure*}

For example, for typical overmassive systems at $z\sim 5$ \citep{Pacucci_2023_JWST}, we obtain ${\cal E}(5) \approx 55 \approx 1.74 \, \mathrm{dex}$ (see Fig. \ref{fig:ratio}).
This indicates that SMBHs in the sample should be $\sim 55$ times overmassive compared to the local \mmstar relation.
Equation \ref{eq:z_evolution} implies that $\Mblack \sim \Mstar$ by $z \sim 30$, in agreement with standard scenarios for the formation of black hole seeds \citep{BL01}.

In Fig. \ref{fig:mmstar}, we use the factor ${\cal E}(z)$ to rescale the local relation \citep{Reines_Volonteri_2015} to higher redshifts. The scaling-up to $z = 5$, the median redshift of the sample of overmassive systems used by \cite{Pacucci_2023_JWST}, agrees remarkably well with the inferred relation determined by the same study.
We also scale up the local relation to $z = 10$ (i.e., a factor of $245$). This scaled-up relation is still too low to explain the extremely overmassive system at $z \sim 10$ described by \cite{Bogdan_2023}. Uncertainties in its black hole and stellar mass could explain the discrepancy.

This redshift evolution of the median value of the \mmstar relation is based on the assumptions that $\Mblack \propto v_c^{5}$ and $M_\star \propto v_c^0$; Section \ref{sec:theory} describes the foundations upon which these assumptions are built. Of course, a shallower relation between black hole mass and circular velocity, and/or a positive, non-zero dependence between the stellar mass and the circular velocity, would lead to a milder redshift evolution, similar to what was found by other studies. For example, \cite{Caplar_2018} derived phenomenologically that $\Mblack/\Mstar \propto (1+z)^{1.5}$. Previous studies (e.g., \citealt{Decarli_2010, Bennert_2011}) also found milder redshift evolutions of the ratio between black hole mass and host stellar masses; for example, the latter study found $\Mblack/\Mstar \propto (1+z)^{1.15}$. Further data at high redshift will clarify the redshift evolution of this fundamental relation.

\subsubsection{Note on the Scatter Around the Relation}
Our model correctly reproduces the redshift evolution of the normalization of the \mmstar relation. Different evolution histories of the single galaxies cause the scatter around the relation; it may be due to second-order effects, such as the specifics of the accretion and merger histories of the single systems, as well as the fact that quasar feedback is likely anisotropic, and the gas distribution non-homogeneous. The prediction of these second-order effects is beyond the scope of this model and requires detailed numerical simulations of the single high-$z$ galactic systems.

\cite{Pacucci_2023_JWST} performed an accurate statistical analysis of the relation between $\Mblack$ and $\Mstar$ for galaxies detected by JWST in the redshift range $4<z<7$. In particular, we adopted the likelihood function defined in \cite{Hogg_2010}, which is appropriate for data characterized by a relation that is ``near-linear but not narrow, so there is an intrinsic width or scatter in the true relationship''.
To account for this intrinsic scatter, one of the three parameters used to describe the inferred relation is $\nu$, which is the orthogonal variance of the best-fit intrinsic scatter. The standard orthogonal scatter in this formalism is $\sqrt{\nu}\sec(\theta)$, where $\theta$ is the angle between the inferred line and the horizontal axis. In the original paper, with 21 data points, the standard scatter is estimated as $0.69$ dex \citep{Pacucci_2023_JWST}.

A new run of the algorithm described in \cite{Pacucci_2023_JWST}, to include all the $35$ galaxies used in the present study (an increase of $67\%$ in data points), led to the following results.
First, the slope and intercept of the new \mmstar relation are consistent with what found in \cite{Pacucci_2023_JWST}: $b = -2.54 \pm 0.75$ (instead of $b = -2.43 \pm 0.83$) and $m = 1.12 \pm 0.08$ (instead of $m = 1.06 \pm 0.09$). The standard orthogonal scatter is $0.53$ dex instead of $0.69$ dex. The scatter decreases because more data points are added in the higher-mass regions of the plot.
Despite adding $67\%$ more data points compared to \cite{Pacucci_2023_JWST}, the intrinsic scatter decreases only by $23\%$.
This example suggests the presence of an intrinsic width or scatter in the true relationship, which is large and due to the single evolutionary histories of the galaxies. Our model aims to characterize the average evolution of the population and cannot describe this scatter.

\section{Discussion and Conclusions}
\label{sec:conclusions}
Before the Hubble Space Telescope performed its first deep field image, it was argued that it would not reveal significantly more galaxies than ground telescopes \citep{Bahcall_1990}.

A similar surprise came during the first year of JWST, which unraveled many galaxies at $z > 4$ hosting a SMBH. Line diagnostics and X-ray detections suggest typical SMBH masses of $\sim 10^6-10^8 \Msun$. With bolometric luminosities $1-2$ orders of magnitude lower than the bright quasars discovered thus far at $z > 6$, their relative faintness allowed the detection of starlight from their hosts. For the first time, observers could investigate the relation between black hole and stellar mass at high-$z$. The data led to the conclusion that high-$z$ SMBHs are $10-100$ times overmassive with respect to the stellar mass of their hosts \citep{Pacucci_2023_JWST}.

Significant uncertainties affect the determination of the stellar mass, derived from SED fitting to galaxy templates, and the SMBH mass, derived from single-epoch virial estimators, based, for example, on the width of the $\Ha$ line of the broad line region (see, e.g., \citealt{Greene_2005}). These methods are calibrated in the local Universe ($z \ll 1$) and have yet to be thoroughly tested at higher redshift (see, e.g., the discussion in \citealt{Maiolino_2023_new}).
Notwithstanding these uncertainties, to retrieve the local scaling relations, the black hole masses (stellar masses) of these high-$z$ overmassive systems would need to be overestimated (underestimated) by a factor $\sim 60$ \citep{Pacucci_2023_JWST}.
Aside from uncertainties on the data side, our model relies on simple, but not simplistic, assumptions that further data and simulations may prove to be incorrect, e.g., specific scaling relations for $\Mblack$ and $\Mstar$ with the circular velocity $v_c$, a specific relation for $v_c = v_c(M_h, z)$, a limited (but realistic) range of Eddington ratios and a fixed fraction $F_q$ of quasar energy trapped by the gas within the galaxy. 

If further data confirms this result, it opens up an important question. Why are these high-$z$ black holes so overmassive with respect to the stellar mass of their hosts while other relations, such as the $\Mblack-\sigma$, hold \citep{Maiolino_2023_new}?

In this Letter, we have developed a model to explain high-$z$ overmassive systems.
The overarching idea is that SMBHs exert an outsized influence on their host galaxies at high-$z$ because their hosts are small, and the black holes have duty cycles close to unity at $z > 4$. Hence, the black hole mass is the primary parameter responsible for high-$z$ star formation quenching. It follows that black hole mass growth is regulated by its energy output, while the stellar mass growth is quenched by it, and its instantaneous value is uncorrelated to the global properties of the host halo.

Our main results are as follows:
\begin{itemize}
    \item In the high-$z$ Universe, the ratio $\Mblack/\Mstar$ controls the average star formation efficiency. If $\Mblack/\Mstar > 8\times 10^{18} (n \Lambda/\fedd)[(\Omega_b M_h)/(\Omega_m M_\star) - 1]$, star formation is quenched by quasar feedback. Once this threshold is crossed, a runaway process brings the originally overmassive system close to the local \mmstar relation.
    \item The local \mmstar relation evolves with redshift as ${\cal E}(z) = \xi(z)^{5/6} (1+z)^{5/2}/\xi(0)^{5/6} \propto (1+z)^{5/2}$, where $\xi(z)$ is weakly dependent on redshift. We find a value of ${\cal E}(z=5) = 55$, which is in excellent agreement with current JWST data and the high-$z$ relation inferred by \cite{Pacucci_2023_JWST}.
\end{itemize}

Our model suggests that early SMBHs, primarily if formed as heavy seeds of initial mass $\sim 10^{4} - 10^5 \Msun$ (see, e.g., \citealt{Ferrara_2014}), affect the evolution of the entire host. Eventually, the activity duty cycle of the quasar drops significantly below unity, and efficient star formation can resume.
Once the galaxy grows via mergers, more stars and cool gas are added. Eventually, stars catch up with the SMBH mass, self-regulation of star formation occurs (see, e.g., \citealt{Wyithe_Loeb_2003}), and the system reaches the local \mmstar relation. Note that the redshift evolution of the \mmstar relation does not prevent the existence of outliers in the mass distribution, both at low and at high redshifts. Our model describes the redshift evolution of the median relation.

Understanding the high-redshift evolution of the scaling relations is fundamental for two reasons. First, it informs us about the physical processes that regulate the growth of the black hole and stellar component (see, e.g., \citealt{Vogelsberger_2014, Schaye_2015, Weinberger_2017, Nelson_2018, Terrazas_2020, Piotrowska_2022, Bluck_2023}). Second, it may inform us of the seeding mechanism that formed the central black hole in the first place. In fact, several studies have shown that a high ratio $\Mblack/\Mstar$ may be indicative of the formation of a heavy seed (see, e.g., \citealt{Agarwal_2013, Natarajan_2017, Visbal_Haiman_2018, Scoggins_2023, Natarajan_2023}) at $z > 20$. The study of the properties of central SMBHs and their hosts at high-$z$, as well as the detection of extremely massive, and rare SMBHs at $z > 10$, will determine if heavy seed formation channels were active in the high-$z$ Universe \citep{Pacucci_2022_search}.

JWST and upcoming facilities such as Euclid, the Rubin Observatory, and the Roman Space Telescope are pushing the observable horizon for black holes farther in redshift and lower in mass. The discovery of still undetected populations of compact objects will ultimately clarify how all the black holes in the Universe formed.

\vspace{10pt}
\noindent \textit{Acknowledgments:} F.P. acknowledges fruitful discussions with Roberto Maiolino and Minghao Yue and support from a Clay Fellowship administered by the Smithsonian Astrophysical Observatory. This work was also supported by the Black Hole Initiative at Harvard University, which is funded by grants from the John Templeton Foundation and the Gordon and Betty Moore Foundation. 

\bibliography{ms}
\bibliographystyle{aasjournal}



\end{document}